\documentclass[10pt,aps,pra,showpacs,amssymb,superscriptaddress,twocolumn,longbibliography]{revtex4-2}

\usepackage[T1]{fontenc}
\usepackage[english]{babel}
\usepackage[utf8]{inputenc}
\usepackage{amsmath,amssymb, amsfonts}
\usepackage[colorlinks=true,citecolor=blue]{hyperref}
\usepackage{graphicx}
\usepackage{svg}
\usepackage{amsmath}
\usepackage{slashed}
\usepackage{bbm}
\usepackage{soul}
\usepackage{array,booktabs}
\usepackage{natbib}
\usepackage{bm}
\usepackage{bbold}
\usepackage{hhline}
\usepackage{xcolor}
\usepackage{mathtools} 
\usepackage{float}
\usepackage{tabularx}
\usepackage{tikz}
\usepackage{physics}
\usepackage[capitalize]{cleveref} 
\usepackage{MnSymbol}
\usepackage{dsfont}
\usepackage{subcaption}
\usepackage[normalem]{ulem}

\usepackage{caption}
\usepackage{tikz}
\usetikzlibrary{quantikz2}
\usepackage{pifont}

\usepackage[table]{xcolor}
\usepackage{booktabs}
\usepackage{multirow}

\def\d(#1,#2,#3){D_{\lambda}(#1|#2#3)}
\def\a(#1,#2){A_{\lambda}\left(#1,#2\right)}
\def\b(#1,#2){B_{\lambda}\left(#1,#2\right)}
\def\som{\sum_{\lambda} q_{\lambda}}
\def\somc{\sum_{x,y} c_{xy}}
\def\ys{\tilde{y}_{x}}
\def\xs{\tilde{x}_{y}}
\def\no{\text{no}}

\newcommand{\id}{\ensuremath{\mathbbm{1}}}
\newcommand{\I}{\mathds{1}}

\def\d(#1,#2,#3){D_{\lambda}(#1|#2#3)}
\def\a(#1,#2){A_{\lambda}\left(#1,#2\right)}
\def\b(#1,#2){B_{\lambda}\left(#1,#2\right)}
\def\som{\sum_{\lambda} q_{\lambda}}
\def\somc{\sum_{x,y} c_{xy}}
\def\ys{\tilde{y}_{x}}
\def\xs{\tilde{x}_{y}}
\def\no{\emptyset}

\begin{document}

\title{Bell and EPR experiments with signalling data}

\author{Lucas Maquedano\href{https://orcid.org/0000-0002-4416-0001}{\includegraphics[scale=0.05]{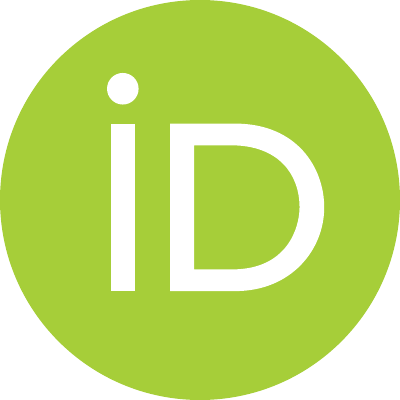}}
}
\thanks{These authors contributed equally.}
\affiliation{Department of Physics, Federal University of Paran\'a, Curitiba, Paran\'a 81531-990, Brazil.}\affiliation{Department of Applied Physics, University of Geneva, Switzerland}

\author{Sophie Egelhaaf \href{https://orcid.org/0009-0006-4769-8161}{\includegraphics[scale=0.05]{orcidid.pdf}}}
\thanks{These authors contributed equally.}
\affiliation{Department of Applied Physics, University of Geneva, Switzerland}

\author{Amro Abou-Hachem}
\affiliation{Physics Department and NanoLund, Lund University, Box 118, 22100 Lund, Sweden}

\author{Jef Pauwels \href{https://orcid.org/0000-0003-4440-7411}{\includegraphics[scale=0.05]{orcidid.pdf}}}
\affiliation{Department of Applied Physics, University of Geneva, Switzerland}\affiliation{Constructor University; 28759 Bremen; Germany}

\author{Armin Tavakoli
\href{https://orcid.org/0000-0001-9136-7411}{\includegraphics[scale=0.05]{orcidid.pdf}}}
\affiliation{Physics Department and NanoLund, Lund University, Box 118, 22100 Lund, Sweden}

\author{Ana C. S. Costa\href{https://orcid.org/0000-0002-4014-0695}{\includegraphics[scale=0.05]{orcidid.pdf}}
}
\affiliation{Department of Physics, Federal University of Paran\'a, Curitiba, Paran\'a 81531-990, Brazil.}

\author{Roope Uola \href{https://orcid.org/0000-0003-3417-4458}{\includegraphics[scale=0.05]{orcidid.pdf}}}
\affiliation{Department of Physics and Astronomy, Uppsala University, Box 516, 751 20 Uppsala, Sweden}
\affiliation{Nordita, KTH Royal Institute of Technology and Stockholm University, Hannes Alfvéns väg 12, 10691 Stockholm, Sweden}

\begin{abstract}
The no-signalling principle is a fundamental assumption in Bell-inequality and quantum-steering experiments. Nonetheless, experimental imperfections can lead to apparent violations beyond those expected from finite-sample statistics. Here, we propose extensions of local hidden variable and local hidden state theories that allow for bounded, operationally quantifiable, amounts of signalling. We show how non-classicality tests can be developed for these models, both through exact methods based on the full set of observed statistics and through corrections to the standard Bell and steering inequalities.  We demonstrate the applicability of these methods via two scenarios that feature apparent signalling: an IBM quantum processor and post-selected data from inefficient detectors.  
\end{abstract}

\maketitle

\section{Introduction}
The seminal Bell and Einstein-Podolsky-Rosen (EPR) scenarios, which study the nature of entanglement, have played a crucial role for fundamental understanding of quantum theory and the development of modern quantum technology. Today, nonlocal correlations are the workhorse behind a number of quantum information tasks, for example, device-independent randomness \cite{Pironio2010}, device-independent quantum key distribution \cite{Zhang2022, Nadlinger2022}, advantages in communication complexity \cite{Buhrmann2010} and self-testing of quantum devices \cite{Supic2020}.

A central assumption behind these tests is the no-signalling principle, which expresses the idea that measurement choices at one location cannot instantaneously influence those at another. In fundamental tests of quantum theory, this principle is justified operationally by enforcing spacelike separation, so that causal influence is excluded by relativity (no influence can travel faster than the speed of light). In contrast, in many practical quantum information settings the no-signalling assumption is taken as a modelling axiom: we rely on our understanding of the devices and of the experimental architecture to assert that no unwanted causal influences exist. In this sense, the assumption becomes part of an idealized theoretical description, rather than something empirically guaranteed.

Nonetheless, many reported Bell and EPR tests display apparent violations of the no-signalling conditions that cannot reasonably be explained by statistical fluctuations alone. This strongly suggests that our idealised models overlook small but systematic influences that are inevitably present in realistic experimental settings. In practice, such deviations are commonly attributed to systematic effects \cite{smania2018}, for example drifts in event rates between different measurement settings, or effective post-selection arising from outcome-dependent detection efficiencies. These issues are  circumvented in fully loophole-free tests \cite{Hensen2015,Zeilinger2015,Shalm2015,Rosenfeld2017}. As a particularly illustrative case, it was shown in \cite{smania2018} that the Bell test currently closest to the quantum bound \cite{Christensen2015} exhibits a violation of the no-signalling constraints exceeding 200 standard deviations.

It is therefore relevant to ask how to meaningfully analyse data from Bell and EPR experiments that feature small amounts of signalling. The aspect of statistical  fluctuations has been addressed \cite{renou2017, Gill2023, engineer2025,cavalcanti2015} but systematic errors, which can lead to violations of the no-signalling condition,  present a distinct issue. 

\begin{figure}
     \centering
     \begin{subfigure}[t]{0.23\textwidth}
         \centering
         \includegraphics[width=.99\textwidth]{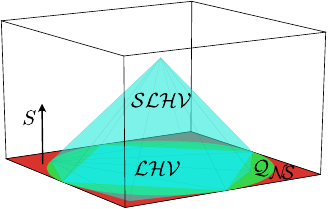}
         \caption{}
         \label{fig:LHV_SPACE}
     \end{subfigure}
     \begin{subfigure}[t]{0.23\textwidth}
         \centering
         \includegraphics[width=.99\textwidth]{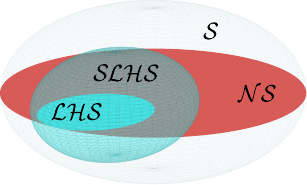}
         \caption{}
         \label{fig:LHS_SPACE}
     \end{subfigure}
     \caption{Geometric illustration of correlations. (a) Bell scenario. The local polytope (LHV) and the quantum set (Q) are subsets of no-signalling probability distributions. When bounded signalling is permitted, the LHV polytope is expanded into the SLHV polytope.  (b) EPR scenario. The unsteerable assemblages (LHS) are a subset of no-signalling assemblages. Bounded signalling inflates the former into the SLHS$_\gamma$ set. Both the SLHV and SLHS$_\gamma$ sets  extend  both within and outside the no-signalling subspace.
     }\label{fig1}
\end{figure}

Here, we take a different approach: rather than post-processing data to enforce no-signalling, we work with explicitly relaxed classical models for both Bell nonlocality and EPR steering. Concretely, we introduce bounded-signalling extensions of local hidden-variable (LHV) and local hidden-state (LHS) models, where the admissible amount of signalling is treated as a parameter that can be estimated directly from the observed data.  This viewpoint is related to prior work that quantifies and accounts for imperfections operationally. Tabia \emph{et al.}~\cite{Tabia2025} use prediction-based methods to detect measurement cross-talk in Bell tests, while Tavakoli \emph{et al.}~\cite{tavakoli2021} study contextual correlations under relaxed assumptions. Most closely in spirit, Vall\'ee \emph{et al.}~\cite{Vallee2024} derive corrected nonclassicality bounds by quantifying the irreducible signalling content of the observed behavior directly at the level of the underlying response functions of the LHV model. Other approaches have considered correlations between local hidden variables and measurement settings \cite{Hall2011, Barrett2011} and possible relaxations of the causal structure \cite{Chaves2015}.

Our approach is both operational and geometric. Operationally, in Bell tests we estimate a ``signalling budget'' directly from the observed changes of local marginals across the other party's settings, while in steering we quantify signalling by how well one party can guess the other's measurement choice from their reduced states. Geometrically, these operational figures of merit specify how far the observed data lie from the no-signalling subspace and, in turn, by how much the classical sets must be inflated: in Bell, this produces an enlarged local polytope, and in steering an enlarged unsteerable set of assemblages; see Fig.~\ref{fig1}.

We first apply these techniques to  Bell inequalities and linear steering witnesses. We then demonstrate the framework on two experimentally relevant scenarios featuring apparent signalling: IBM quantum hardware, where residual signalling can lead to false-positives for steering, and Bell tests with post-selection under inefficient detectors, where a naive analysis can spuriously suggest post-quantum violations unless the signalling is explicitly taken into account.

\section{Bell tests with signalling} Consider a Bell experiment in which Alice and Bob independently select inputs $x$ and $y$, respectively, and their respective outcomes are denoted $a$ and $b$. In a faithful implementation, the measurement events are independent and therefore the joint probability distribution $p(a,b\lvert x,y)$  satisfies  no-signalling, namely 
\begin{equation}
\sum_a p(a,b\lvert x,y)=p(b\lvert y), \quad \sum_b p(a,b\lvert x,y)=p(a\lvert x).
\end{equation}

We are interested in situations where the experimental data $p_\text{exp}(a,b\lvert x,y)$ does not comply with this principle. To test for nonlocality in data that features signalling, we must extend local hidden variable (LHV) models with the capability to signal. To this end, we consider the signalling LHV (SLHV) model 
\begin{equation}\label{SLHV}
p(a,b|x,y)=\sum_{\lambda} q_{\lambda} D^A_{\lambda}(a|x,y) D^B_{\lambda} (b|x,y)
\end{equation}
where the hidden variable $\lambda$, which has some distribution $q_\lambda$, is used to map both $x$ and $y$ into the outputs of the observers via the deterministic assignment $D^A_\lambda$ for Alice and $D^B_\lambda$ for Bob. This contrasts a standard LHV model, for which no-signalling would have implied that $D^A_{\lambda}(a|x,y)=D^A_{\lambda}(a|x)$ and  similarly for Bob. In contrast, the more general model \eqref{SLHV} can reproduce any probability distribution $p(a,b\lvert x,y)$ unless the hidden variable is ontologically constrained. Therefore, we consider the restriction that the average ontological amount of signalling is bounded as 
\begin{align}\nonumber
&\sum_{\lambda} 	q_{\lambda} |D^A_{\lambda}(a|x,y)-D^A_{\lambda}(a|x,y') | \leq \alpha^{ax}_{yy'}, \\
&\sum_{\lambda} 	q_{\lambda} |D^B_{\lambda}(b|x,y)-D^B_{\lambda}(b|x',y) | \leq \beta^{by}_{xx'},
\end{align} 
for Alice and Bob respectively. One may think of the first condition as $\alpha^{ax}_{yy'}(\lambda)\equiv |D^A_{\lambda}(a|x,y)-D^A_{\lambda}(a|x,y') |$ being the table of ``signalling costs'' for the response function $D^A_\lambda$ and that the average signalling cost, namely $\sum_\lambda q_\lambda \alpha^{ax}_{yy'}(\lambda)$ is bounded by a set of parameters $\alpha^{ax}_{yy'}$ that we are free to choose. The analogue applies to the second condition. 
This approach is operationally motivated, because possible values for the  signalling parameters $\alpha^{ax}_{yy'}$ and $\beta^{by}_{xx'}$ can be estimated directly from experimental data. One appealing choice is to take $\alpha^{ax}_{yy'}= \rvert p_\text{exp}(a\lvert x,y)-p_\text{exp}(a\lvert x,y')\lvert$ and $\beta^{by}_{xx'}=\rvert p_\text{exp}(b\lvert x,y)-p_\text{exp}(b\lvert x',y)\lvert$.  Note that if all $\alpha^{ax}_{yy'}$ and $\beta^{by}_{xx'}$ are zero, the standard LHV model is recovered.

\subsection{Linear programming for SLHV models}
Like the set of correlations achievable by LHV models, the set of correlations achieved by an SLHV model is geometrically represented as a polytope. For given   $\alpha^{ax}_{yy'}$ and $\beta^{by}_{xx'}$, the SLHV polytope can be viewed as an expansion of the LHV polytope  in both the signalling and the no-signalling part of the space of probability distributions; see Fig.~\ref{fig:LHV_SPACE}. 

As a consequence of this polytope structure, one can decide whether $p(a,b|x,y)$ belongs to the SLHV polytope by means of a linear program which can be formulated as follows,

	\begin{equation}
	\begin{aligned}
		&\max _{ \{ q_{\lambda} \} } &&\quad v \\
		&\text{s.t.} &&  v\ p(a,b\lvert x,y) + \frac{1-v}{n_{A}n_{B}} = \sum_{\lambda} q_{\lambda} D^{A}_{\lambda}(a|x,y)D^{B}_{\lambda}(b|x,y) \\
		&  && \sum_{\lambda} q_{\lambda}=1,\\ 
		& && q_{\lambda}\geq 0 \quad \forall \lambda, \\
		& && \sum_{\lambda} q_{\lambda} |D^{A}_{\lambda}(a|x,y) -D^{A}_{\lambda}(a|x,y')|\leq \alpha^{ax}_{yy'}  \\
		& && \sum_{\lambda} q_{\lambda} |D^{B}_{\lambda}(b|x,y) -D^{B}_{\lambda}(b|x',y)| \leq \beta^{by}_{xx'} .\\
	\end{aligned}
	\label{eq:visibility_linear_program}
\end{equation}
Here, we label the number of outcomes for Alice and Bob by $n_A$ and $n_B$ and introduce a mixture between $p(a,b|x,y)$ and the uniform distribution, with a signal strength $v\in[0,1]$. We can think of $v$ as a quantifier how well the SLHV model can approximate $p(a,b|x,y)$. 

If no SLHV model exists (i.e.~$v<1$), one can in addition leverage the duality theorem of linear programming to determine an appropriate signalling Bell inequality that is violated by $p_\text{exp}(a,b|x,y)$. This provides a general method to test whether signalling data in Bell tests is compatible with an SLHV model. The linear programming dual is detailed in Appendix~\ref{section:Visibility_dual} and we also provide an implementation which is available at \cite{LP_github}.

\subsection{Correcting Bell inequalities} 
The linear programming method has to be applied on a case-to-case basis and can  become expensive when used beyond the few smallest Bell scenarios. We therefore show how to analytically update the bound of standard Bell inequalities so that they apply also to SLHV models. 

To this end, consider that we have the full correlation dichotomic Bell inequality
\begin{equation}
\mathcal{W}=\sum_{x,y} c_{xy} E_{xy} \leq W_\text{LHV},
\label{eq:LHV_inequality}
\end{equation}
where $c_{xy}\in \mathbbm{R}$, $E_{xy}=\sum_{a,b=0,1}(-1)^{a+b}p(a,b|x,y)$ is the joint expectation value and $W_\text{LHV}$ is the standard LHV bound. In Appendix~\ref{section:Dichotomic_bell_correction} we prove that for given signalling parameters $\alpha^{ax}_{yy'}$ and $\beta^{by}_{xx'}$, the SLHV bound of the same Bell parameter becomes 
\begin{equation}
\mathcal{W}\leq W_{\text{LHV}} +  \sum_{x,y} |c_{xy}|\min_{u_x,v_y} \left(\sum_{a} \alpha^{ax}_{yu_x}+\sum_{b}\beta^{by}_{xv_y}\right).
\end{equation}
For example, the most commonly considered Bell test is that of  Clauser-Horne-Shimony-Holt (CHSH) for which the above simplifies to 
\begin{equation}
    \mathcal{W}_\text{CHSH} \leq 2 +  2\sum_{x} \alpha^{0x}_{01}+2\sum_{y} \beta^{0y}_{01}.
\end{equation}
We see that when correcting Bell inequalities for signalling, the relevant quantity is the sum-total of the signalling parameters.

\section{EPR tests with signalling}
In a steering scenario, the measurements performed by Bob are assumed to be fully characterised. Consequently, the relevant objects describing quantum correlations are assemblages,
\[
\sigma_{a|x} = \mathrm{tr}_A \!\left[(M_{a|x} \otimes \openone_B)\rho^{AB}\right],
\]
which are prepared remotely for Bob by Alice’s local measurements $\{M_{a|x}\}$ on a shared state $\rho^{AB}$. The no-signalling principle then imposes the condition
\begin{equation}\label{NS_condition}
    \sum_a \sigma_{a|x} := \mu_x = \mu \hspace{10pt} \forall x \,.
\end{equation}

In analogy with our discussion of Bell nonlocality, we are interested in steering assemblages $\{\sigma_{a\lvert x}\}$ that violate this no-signalling condition. An operationally meaningful way to quantify deviations from the no-signalling principle is by asking how well the classical label $x$ can be discriminated when encoded into the states $\{\mu_x\}_x$. This corresponds to  the guessing probability 
\begin{equation}\label{Eq:GuessProb}
        P_g(\{\mu_x\}_x) := \frac{1}{m_A}\max_{\{N_x\}} \sum_{x}  \tr(N_x \mu_x),
\end{equation}
 where $m_A$ is the number of settings for Alice and $\{N_x\}$ is a quantum measurement. This quantity is computable by a semidefinite program \cite{Bae2015}. When no-signalling is respected, one has $P_g=\frac{1}{m_A}$ and otherwise  $P_g>\frac{1}{m_A}$. In addition to being a natural quantifier of accessible information \cite{tavakoli2020, tavakoli2022} that also has been used to quantify systematic deviations in contextuality experiments \cite{tavakoli2021}, the guessing probability admits a relevant geometric interpretation as a robustness measure for the set of assemblages that respect no-signalling \cite{tavakoli2021}.

With $P_g$ as our signalling parameter, we must extend the conventional LHS models with the capability to signal. We consider the signalling LHS (SLHS$_\gamma$) model 
\begin{equation}\label{Eq:LHSsignalling}
\sigma_{a\lvert x}=\sum_\lambda q_\lambda D^A_\lambda(a\lvert x) \sigma_{\lambda,x},
\end{equation}
where $\sigma_{\lambda,x}$ is a state and $q_\lambda$ is a probability distribution over $\lambda$. Note that the usual LHS model corresponds to substituting $\sigma_{\lambda,x}$ for $\sigma_{\lambda}$, and that the former enables the signalling. Naturally, any assemblage admits a model of the above form. In analogy with the case of Bell scenarios, we consider that the ontological value of the signalling parameter is bounded as 
\begin{equation}\label{SLHS}
\sum_\lambda q_\lambda P_g\left(\{\sigma_{\lambda,x}\}_x\right)\leq \gamma
\end{equation}
A relevant  choice of $\gamma$ is to select it as the guessing probability of the experimentally observed assemblage, i.e.~$\gamma=P_g\left(\{\sigma^\text{exp}_{x}\}_x\right)$.

\subsection{Semidefinite programming for SLHS$_\gamma$ models}
For no-signalling assemblages, steerability can be decided efficiently via semidefinite programming (SDP)~\cite{cavalcanti2016}. We find that an analogous SDP can be  formulated also in the presence of signalling.  To arrive at this formulation, we first write  the guessing probability as the SDP
\begin{align}\nonumber
P_g(\{\sigma_x\}_x) =& \max_{\{N_x\}} \quad\frac{1}{m_A}\sum_x \tr\left(N_x\sigma_x\right) \\
& \text{s.t} \quad N_x\succeq0 , \qquad \sum_x N_x=\openone.
\end{align}
Since strong duality holds, we can express it as the dual program
\begin{align}\nonumber
P_g(\{\sigma_x\}_x) =& \min_{Z} \quad \tr(Z) \\
& \text{s.t} \quad Z\succeq \frac{1}{m_A} \sigma_x.
\end{align}
Consequently, our signalling parameter can be written as 
\begin{align}\nonumber
\sum_\lambda q_\lambda P_g\left(\{\sigma_{\lambda,x}\}_x\right) \quad = \quad & \min_{Z_\lambda} \quad \sum_\lambda \tr(Z_\lambda) \\
& \text{s.t} \quad Z_ \lambda\succeq \frac{1}{m_A} \tilde{\sigma}_{\lambda,x},
\end{align}
where we have defined $\tilde{\sigma}_{\lambda,x}=q_\lambda \sigma_{\lambda,x}$.
We can bound the signalling parameter by $\gamma$ by imposing this limit on the objective of this program. This lets us express the membership problem to the SLHS$_\gamma$ set as the SDP
\begin{equation}
\begin{aligned}
    \text{find} \quad & \{\tilde{\sigma}_{a|x}\}_{a,x}\\
 \text{s.t.} & \quad  \sigma_{a|x}^{\operatorname{SLHS}_\gamma} = \sum_\lambda D(a|x,\lambda)\tilde{\sigma}_{\lambda, x}\\
& \sum_\lambda \tr(Z_\lambda) \leq \gamma\\
& Z_\lambda \succeq \frac{1}{m_A} \tilde{\sigma}_{\lambda,x} \quad \forall \lambda, x
\end{aligned}
\end{equation}
Moreover, by standard SDP duality, one can extract an optimal SLHS$_\gamma$ witness (inequality) tailored to the tested assemblage.

It is convenient to define a distance measure with respect to the set SLHS$_\gamma$. For this, we consider the robustness measure. This is defined as 
    \begin{align}\label{SDPmembership}
        SR_{\text{SLHS}_\gamma}\qty(\{\sigma_{a|x} \}) = & \min r\ge 0 \\
        & \text{s.t. } \frac{\sigma_{a|x} + r \tau_{a|x}}{1+r} \in \text{SLSH}_\gamma \quad \forall a,x, \nonumber
    \end{align}
    where the minimum is taken over all  (possibly signalling) $\{\tau_{a|x}\}$ assemblages. Like the membership problem, the robustness can be computed as an SDP by a small modification of \eqref{SDPmembership}; see Appendix~\ref{app:sdpsteering}. A numerical implementation is available at~\cite{maquedano2025}.

\subsection{Corrections to steering inequalities}

It is possible to geometrically view the SLHS$_\gamma$ model as a robustness measure. Specifically, we show in Appendix~\ref{app:LHSgeom} that any model that obeys Eqs.~\eqref{Eq:LHSsignalling} and \eqref{SLHS} admits  the following decomposition
\begin{equation}\label{Eq:LHS_Sr}
      \sigma_{a|x}= (1+t) \sigma_{a|x}^{\text{LHS}} - t \tau_{a|x}.
\end{equation}
where  $\{\tau_{a\lvert x}\}$ is a signalling assemblage ($\tau_{a\lvert x}\succeq 0$ and $\sum_a \tr(\tau_{a\lvert x})=1$),  $\{\sigma_{a|x}^{\text{LHS}}\}$ is an LHS assemblage and $0\leq t\leq m_A \gamma$. Here, we may think of $t$ as a quantifier of the contribution from the space of signalling assemblages.  

Thanks to this geometric interpretation of SLHS$_\gamma$ models, it is also possible to obtain signalling corrections to standard steering inequalities. Any steering inequality can be written in the form
\begin{equation}\label{Eq:LHSbound}
    \mathcal{L}=\sum_{a,x} \tr\!\left(\sigma_{a|x} W_{a|x} \right) \leq L_\text{LHS},
\end{equation}
where $W_{a\lvert x}$ are positive semidefinite operators, and $L_\text{LHS}$ is respected by all LHS assemblages. To obtain a bound that holds for any assemblage admitting an SLHS$_\gamma$ model, we use Eq.~\eqref{Eq:LHS_Sr} with $t\le m_A\gamma$ to get
\begin{align}\label{Eq:LinearInequalityLinearAdjustment}\nonumber
     \mathcal{L}
     &= (1+t)\sum_{a,x}\tr\!\left(\sigma_{a|x}^{\mathrm{LHS}} W_{a|x}\right)
        - t\sum_{a,x}\tr\!\left(\tau_{a|x} W_{a|x}\right)\\
    &\le (1+t)L_\text{LHS}
    \;\le\; L_\text{LHS}(1+m_A \gamma).
\end{align}
While this bound is straightforward, it is not expected to be tight. We note that the adjustment technique can also be applied to correct tests of entanglement dimensionality in steering \cite{designolle2021}, see Appendix~\ref{app:SN}.

\section{Applications}
We discuss two different scenarios in which apparent signalling arises in experimental data. For each, we use  the proposed methods to analyse the experiment for steering and nonlocality.

\subsection{Certifying entanglement on IBM chip} 
Consider the certification of  steering on a superconducting quantum processor. In this scenario, the no-signalling condition is not exactly satisfied and must therefore be accounted for explicitly.
The protocol uses a quantum circuit that first prepares a two-qubit isotropic state $\rho_{v}$. The untrusted party then implements the three Pauli measurements, while the trusted party performs quantum state tomography to estimate the assemblage $\{\sigma_{a|x}\}_{a,x}$. The complete circuit implementation is described in~\cite{maquedano2025}.

Our goal is to investigate the amount of signalling and steering present in the resulting assemblages. To distinguish the effects of finite statistics from those of residual physical signalling, we vary the number of initial state preparations $N$ used to reconstruct the assemblage. In particular, for small $N$, the observed amount of signalling can fluctuate significantly across different experimental runs.

\renewcommand{\arraystretch}{1.2} 
\setlength{\lightrulewidth}{0.03em}
\begin{table}[h]
    \centering
    \begin{tabular}{ccccccc} 
        \toprule
        & \multicolumn{2}{c}{$P_{\text{g}}$} & \multicolumn{2}{c}{$SR_{\text{SLHS}_\gamma}$} & \multicolumn{2}{c}{$SR_{\text{SLHS}_\gamma}^{\I}$} \\ 
        \cmidrule(lr){2-3} \cmidrule(lr){4-5} \cmidrule(lr){6-7}
        $N$ & $v = 0.8$ & $v = 1.0$ & $v = 0.8$ & $v = 1.0$ & $v = 0.8$ & $v = 1.0$ \\
        \midrule
        10    & 0.5040   & 0.4468 & 0.0264 & 0.1694 & 0.0776 & 0.4226 \\
        600   & 0.3509   & 0.3721 & 0.0158 & 0.1325 & 0.0603 & 0.3477 \\
        2500  & 0.3599   & 0.3563 & 0.0292 & 0.1249 & 0.0905 & 0.3458 \\
        10000 & 0.3748   & 0.3645 & 0.0499 & 0.1490 & 0.1605 & 0.4055 \\
        \bottomrule
    \end{tabular}
    \caption{Experimental results obtained from the IBM quantum processor. The table presents the guessing probability ($P_{\text{g}}$) and steering signalling robustness ($SR_{\text{sig}}$) for isotropic states $\rho_v$ with visibilities $v=0.8$ and $v=1.0$. The values are shown as a function of the number of preparations $N$ used for the assemblage tomography. Despite the presence of a significant amount of signalling, we can still certify steering, even for small sample sizes.}
     \label{tab:experimental_data}
\end{table}

For visibilities $v=0.8$ and $v=1$, we analyse both the guessing probability and the steering signalling robustness as a function of $N$ using experiments on the IBM quantum platform. The circuit was executed on the \texttt{ibm\_brisbane} backend, equipped with an Eagle r3 family processor comprising 127 superconducting transmon qubits. This device features a heavy-hexagonal topology, in which each qubit interacts with two or three neighbors~\cite{ghanem2025, brisbane_details}.

As $N$ increases, purely statistical fluctuations are expected to induce apparent signalling that decreases approximately as $1/\sqrt{N}$. However, the experimental data shown in Table~\ref{tab:experimental_data} exhibit a clear saturation at a nonzero plateau for large $N$. This behaviour indicates the presence of systematic sources of signalling, plausibly arising from residual hardware imperfections~\cite{Krantz2019,Kjaergaard2020,Ni2022}.

Given the guessing probability for each experimentally reconstructed assemblage, we compute the corresponding white noise and generalized robustness. The results are summarised in Table~\ref{tab:experimental_data}. Note that no signalling corresponds to $P_g = 1/3$, while the theoretical maximum amount of signalling is $P_g = 2/3$~\cite{Holevo1973}. The observed values therefore indicate substantial deviations from the no-signalling ideal.

Despite this residual signalling, steering can still be certified already for modest sample sizes. For $v=0.8$, the state is less steerable and the circuit implementation is more complex, leading to stronger residual signalling.

\subsection{Post-selection in Bell experiments}

\begin{figure}
    \centering
    \includegraphics[width=1\linewidth]{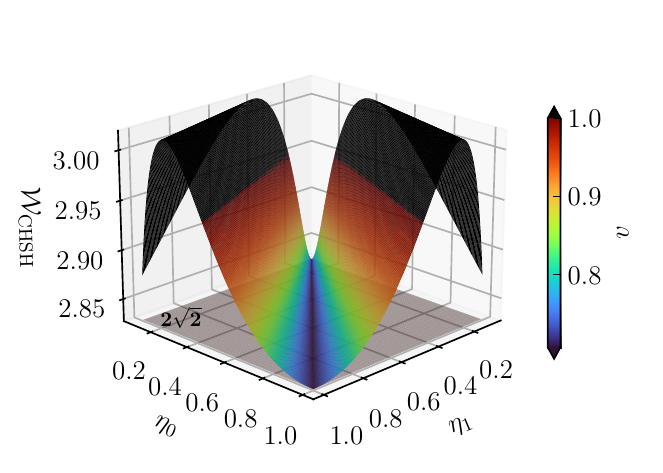}
    \caption{Standard quantum CHSH test but when post-selecting on successful detections with outcome efficiencies $\eta_{0}$ and $\eta_{1}$. The heat map represents how well the statistics can be approximated by an SLHV model, quantified through the visibility parameter $v$.}\label{fig_efficiency}
\end{figure}

A  relevant way in which apparent signalling can arise in Bell experiment is post-selection. To that end, consider that the estimation of $p(a,b\lvert x,y)$ is based only on the events that are detected, and that these events have a probability $\eta_{a,x}^A$ and $\eta_{b,y}^B$ of being registered by the detection devices of Alice and Bob respectively. The efficiency can be estimated as 
\begin{equation}
	\eta^A_{a|x}= \frac{n_{x}(a)}{n_{x}(a)+n_{x}(\no)}, 
\end{equation}
where $n_{x}(a)$ is the number of events with outcome $a$ and input $x$ and $n_x(\no)$ is the number of no-click events. Since there is effectively one extra outcome the  alphabet becomes $\tilde{a}\in[n_A]\cup \no$. We  similarly  have $\tilde{b}\in[n_A]\cup \no$ for Bob and associate detection efficiencies $\eta^B_{b\lvert y}$. The POVMs associated to the inefficient measurements become
\begin{align}
	&A_{\tilde{a}|x} = \begin{cases}
		\eta_{a|x}^A A_{a|x} & \text{if } \tilde{a}=a, \\
		\openone - \sum_{a} \eta_{a|x}^A A_{a|x} &  \text{if } \tilde{a}=\no,
	\end{cases}\\
  &	B_{\tilde{b}|y} = \begin{cases}
		\eta_{b|y}^B B_{b|y} & \text{if } \tilde{b}=b, \\
		\openone - \sum_{b} \eta_{b|y}^B B_{b|y} &  \text{if } \tilde{b}=\no
	\end{cases},
\end{align}
where  $\{ A_{a|x}\}$ and $\{ B_{b|y} \}$ are the POVMs of Alice and Bob with unit efficiency. 
Consider now that when Alice and Bob perform their measurements they post-select on the successful detection events, i.e.~when $\tilde{a}\in[n_A]$ and $\tilde{b}\in [n_B]$. The post-selected probability distribution becomes 
\begin{align}\nonumber
	p(a,b|x,y)&=\frac{p(\tilde{a}=a ,\tilde{b}=b|x,y)}{N_{x,y}} \\
    & = \frac{\eta_{a|x}^A \eta_{b|y}^B}{N_{x,y}}\tr( \left(A_{a|x}\otimes B_{b|y} \right)\rho),
\end{align}
where we have defined $N_{x,y}=\sum_{\tilde{a} \neq \no , \tilde{b} \neq \no} p(\tilde{a},\tilde{b}|x,y)$ as the sum-total probability of both Alice and Bob obtaining a successful event.

For simplicity of presentation, we consider only that the different outcomes have different efficiencies, namely $\eta_0$ and $\eta_1$. A consequence of such post-selection is that one may observe not only violations of no-signalling but also a CHSH parameter that exceeds the  Tsirelson bound $\mathcal{W}_\text{CHSH}\leq 2\sqrt{2}$.  How much Bell nonlocality is actually present in such a post-selected quantum distribution once we take the signalling parameters into account?  We have considered the standard quantum strategy for reaching Tsirelson's bound, namely by sharing a maximally entangled two-qubit state and performing local anticommuting measurements, but post-selected on successful detection events with efficiencies $\eta_0$ and $\eta_1$. This gives the CHSH parameter
\begin{equation}
\mathcal{W}_\text{CHSH}=\frac{2 \left(\eta_0^4+16 \sqrt{2} \eta_1 \eta_0^3-2 \eta_1^2 \eta_0^2+16 \sqrt{2}
   \eta_1^3 \eta_0+\eta_1^4\right)}{\eta_0^4+12 \eta_1 \eta_0^3+6 \eta_1^2
   \eta_0^2+12 \eta_1^3 \eta_0+\eta_1^4}.
\end{equation}
Note that for the special case of $\eta_0=\eta_1$ this becomes $\mathcal{W}_\text{CHSH}=2\sqrt{2}$ but values $\mathcal{W}_\text{CHSH}>2\sqrt{2}$ are possible. 

In Fig.~\ref{fig_efficiency} we plot the CHSH parameter and illustrate how it can  exceed Tsirelson's bound for  $\eta_0\neq \eta_1$. For every such post-quantum Bell inequality violation, we have computed the associated signalling parameters (see Appendix ~\ref{app:postselection}) and used the linear programming in Eq.~\eqref{eq:visibility_linear_program} to determine its distance to the SLHV polytope through the largest  $v\in[0,1]$ compatible with an SLHV model. The result is displayed as a heat map in Fig.~\ref{fig_efficiency}. In the black regions, there is no nonlocality in spite of the post-quantum CHSH parameter. In the coloured region, some nonlocality still survives, but the most noise-robust results are found along the no-signalling corridor corresponding to $\eta_0=\eta_1$.

\section{Conclusions} In this work, we developed an operational and geometric framework to analyse Bell and EPR experiments in which the observed data exhibit small but non-negligible signalling. Rather than post-processing the data to enforce no-signalling, we introduce bounded-signalling extensions of LHV and LHS models, leading to enlarged classical sets that remain testable by convex criteria.

On the Bell side, we characterised bounded-signalling LHV models through experimentally accessible signalling ``budgets'' and showed that compatibility with the resulting bounded-signalling local polytope can be decided by linear programming. This also provides, via duality, optimal Bell inequalities tailored to a given dataset, as well as analytic corrections to standard Bell inequalities that account for the allowed amount of signalling. We illustrated these methods in a post-selection scenario with inefficient detectors, where apparent signalling can spuriously enhance standard Bell parameters if not explicitly accounted for.

On the steering side, our method is based on quantifying signalling via the guessing probability of one party's measurement choice from the other's reduced states. This quantity is dual to a geometric signalling robustness, which we use to translate the observed signalling into an inflation of the unsteerable set of assemblages. We demonstrated the applicability of this approach by deriving adjustments for linear steering inequalities, including genuine high-dimensional steering inequalities, and by developing an SDP method to construct optimal steering witnesses for a given signalling assemblage. The latter was demonstrated on data obtained from a multi-qubit circuit implemented in Qiskit and executed on IBM quantum hardware.

Our work underlines the need to account for apparent signalling appearing in experimental data collected in Bell and steering experiments. However, as we has seen, this is only possible by introducing additional assumptions on the ontology of the experiment. Such assumptions cannot be grounded in a fundamental principle or theory, and are therefore not on par with the fundamental nature of a Bell test and its device-independent applications in quantum information. For that, one requires tests that rigorously satisfy the requirements for no-signalling.

\begin{acknowledgements}
This work has been supported by the Swedish Research Council (grant no. 2024-05341), The Wallenberg Initiative on Networks and Quantum Information (WINQ), the Swiss National Science Foundation (Ambizione PZ00P2- 202179 and SwissMAP). A.C.S.C. was supported by CNPq/Brazil, Grants No. 308730/2023-2 and No. 409673/2022-6. L.M. was supported by CAPES/PRINT (grant No. 88887.935392/2024-00) and CAPES/PROEX (grants No. 88887.695811/2022-00 and No. 88887.000763/2024-00). A.T.~is supported by the Knut and Alice Wallenberg Foundation through the Wallenberg Center for Quantum Technology (WACQT), the Swedish Research Council under Contract No.~2023-03498 and the Swedish Foundation for Strategic Research. A.A.H.~was supported by the EDU-WACQT program funded by Marianne and Marcus Wallenberg Foundation.
\end{acknowledgements}

\bibliography{references}

\newpage
\appendix

\onecolumngrid

\section{Linear programming methods for SLHV}\label{section:Visibility_dual}

Consider a Bell scenario in which Alice and Bob independently select inputs $x\in[m_A]$ and $y\in[m_B]$, respectively, where $[m]=\{0,\ldots,m-1\}$. Their respective outcomes are denoted $a\in[n_A]$ and $b\in [n_B]$. The probability distribution is denoted $p(a,b\lvert x,y)$. We consider a noisy form of this distribution, which we define as $vp(a,b\lvert x,y)+\frac{1-v}{n_An_B}$, where  $\frac{1}{n_An_B}$ is the uniform distribution and $v\in[0,1]$ is the visibility parameter. We can use $v$ as a quantifier of the distance between $p(a,b\lvert x,y)$ and the SLHV polytope. This distance can be computed as the following linear program. 

	\begin{equation}
	\begin{aligned}
		&\max _{ \{ q_{\lambda} \} } &&\quad v \\
		&\text{s.t.} && \quad v\ p(a,b\lvert x,y) + \frac{1-v}{n_{A}n_{B}} = \sum_{\lambda} q_{\lambda} D^{A}_{\lambda}(a|x,y)D^{B}_{\lambda}(b|x,y) \quad \forall a,b,x,y, \\
		&  && \sum_{\lambda} q_{\lambda}=1,\\ 
		& && q_{\lambda}\geq 0 \quad \forall \lambda, \\
		& && \sum_{\lambda} q_{\lambda} |D^{A}_{\lambda}(a|x,y) -D^{A}_{\lambda}(a|x,y')|\leq \alpha^{ax}_{yy'} \quad \forall a,x,y,y', \\
		& && \sum_{\lambda} q_{\lambda} |D^{B}_{\lambda}(b|x,y) -D^{B}_{\lambda}(b|x',y)| \leq \beta^{by}_{xx'} \quad \forall b,y,x,x'.\\
	\end{aligned}
\end{equation}

By the duality theorem of linear programming, we can compute the linear program dual to the above and conserve the solution, i.e.~the distance to the SHLV polytope. This dual program takes the form 
	\begin{equation}
	\begin{aligned}
		&\min_{\substack{
				\{c_{abxy}\}, \mu, \\
				\{d_{axyy'}\},\{ e_{byxx'}\}
		}}&& 1-\mu +\sum_{a,b,x,y} c_{abxy} p(a,b|x,y)+ \sum_{a,x,y,y'} d_{axyy'} \alpha^{ax}_{yy'} + \sum_{b,y,x,x'} e_{byxx'} \beta^{by}_{xx'} \\
		&\text{s.t.} && \\
		&&& 1 + \sum_{a,b,x,y} c_{abxy} \left(p(a,b\lvert x,y) - \frac{1}{n_{A}n_{B}}\right) = 0, \\ 
		&&& \sum_{a,b,x,y} c_{abxy} D^{A}_{\lambda}(a|x,y) D^{B}_{\lambda}(b|x,y)   \quad + \sum_{a,x,y,y'} d_{axyy'} R^{ax}_{yy'\lambda} + \sum_{b,y,x,x'} e_{byxx'} T^{by}_{xx'\lambda} \geq \mu \quad \forall \lambda ,  \\
		&&& d_{axyy'} \geq 0 \quad \forall a,x,y,y', \\
		&&& e_{byxx'} \geq 0 \quad  \forall b,y,x,x',
	\end{aligned}
\end{equation}
where  $R^{ax}_{yy'\lambda}=|D^{A}_{\lambda}(a|x,y)-D^{A}_{\lambda}(a|x,y')|$ and $T^{by}_{xx'\lambda}=|D^{B}_{\lambda}(b|x,y)-D^{B}_{\lambda}(b|x',y)|$. 

In case $p(a,b\lvert x,y)$ is not a member of the SLHV polytope, i.e~the solution is $v<1$, then the dual program allows us to extract a signalling Bell inequality that detects $p(a,b\lvert x,y)$. This inequality can be extracted from the dual by summing the second constraint above under the weight $q_\lambda$. That gives
\begin{equation}
    \sum_{a,b,x,y} c_{abxy} p(a,b|x,y)\geq \mu  -\sum_{a,x,y,y'} d_{axyy'} \alpha^{ax}_{yy'} -\sum_{b,y,x,x'} e_{byxx'} \beta^{by}_{xx'} 
\end{equation}
for all SLHV models.

\section{signalling correction to dichotomic Bell inequalities \label{section:Dichotomic_bell_correction}}

Consider the full correlation dichotomic Bell inequality
\begin{equation}
\mathcal{W}=\sum_{x,y} c_{xy} E_{xy} \leq W_\text{LHV},
\end{equation}
where $c_{xy}$ are arbitrary real numbers and $E_{xy}=\sum_{a,b} (-1)^{a+b} p(a,b|x,y)$. 

We now show how to update the bound for an SLHV model with signalling parameters $\alpha^{ax}_{yy'}$ and $\beta^{by}_{xx'}$.
In an SLHV model the correlators take the form 
 \begin{equation}
 	\begin{aligned}
 	E_{xy}&= \sum_{a,b=0,1} (-1)^{a+b} \sum_{\lambda}  q_{\lambda}D^{A}_{\lambda}(a|x,y) D^{B}_{\lambda}(b|x,y)\\
 	&=\sum_{\lambda} q_{\lambda} \left[ \sum_{a=0,1} (-1)^{a} D^{A}_{\lambda}(a|x,y) \right]\left[ \sum_{b=0,1} (-1)^{b}D^{B}_{\lambda}(b|x,y) \right] = \sum_{\lambda} q_{\lambda}A_{\lambda}(x,y) B_{\lambda}(x,y),
\end{aligned}
 \end{equation}
 where we defined $A_{\lambda}(x,y)=\sum_{a=0,1} (-1)^{a} D^{A}_{\lambda}(a|x,y)$ and $B_{\lambda}(x,y)=\sum_{b=0,1} (-1)^{b}D^{B}_{\lambda}(b|x,y)$.  With this in hand, we can derive the following inequality
 \begin{equation}
 \begin{aligned}
 |\sum_{\lambda}q_{\lambda} \b(x,y) (A_{\lambda}(x,y)-A_{\lambda}(x,y'))| & \leq  \sum_{\lambda} q_{\lambda} |\b(x,y)||A_{\lambda}(x,y)-A_{\lambda}(x,y')|\\
 &= \sum_{\lambda} q_{\lambda}  |\sum_{a} (-1)^{a} (D^{A}_{\lambda}(a|x,y)-D^{A}_{\lambda}(a|x,y'))|\\
 &\leq \sum_{a} \sum_{\lambda} q_{\lambda}  |D^{A}_{\lambda}(a|x,y)-D^{A}_{\lambda}(a|x,y')|\leq  \sum_{a}  \alpha^{ax}_{yy'},
 \end{aligned}
 \end{equation} 
where we used that $|\b(x,y)|=1$, the triangle inequality and the definition of the SLHV model. Analogously for Bob we have 
\begin{equation}
 \begin{aligned}
 |\sum_{\lambda}q_{\lambda} \a(x,y) (\b(x,y) -\b(x',y)| & \leq  \sum_{\lambda} q_{\lambda} |\a(x,y)||\b(x,y) - \b(x',y)|\\
 &= \sum_{\lambda} q_{\lambda}  |\sum_{b} (-1)^{b} (D^{B}_{\lambda}(b|x,y)-D^{B}_{\lambda}(b|x',y))|\\
 &\leq \sum_{b} \sum_{\lambda} q_{\lambda}  |D^{B}_{\lambda}(b|x,y)-D^{B}_{\lambda}(b|x',y)|\leq \sum_{b}  \beta^{by}_{xx'}.
 \end{aligned}
 \end{equation}
The relations can equivalently be written as 
\begin{align}\label{eq:inequality_inA}
&-\sum_{a}  \alpha^{ax}_{yy'} \leq \sum_{\lambda}q_{\lambda } \b(x,y)\left(A_{\lambda}(x,y)-A_{\lambda}(x,y') \right) \leq \sum_{a}\alpha^{ax}_{yy'},\\
& -\sum_{b}  \beta^{by}_{xx'} \leq \sum_{\lambda} q_{\lambda} \a(x,y)\left( B_{\lambda}(x,y)-B_{\lambda}(x',y) \right) \leq \sum_{b} \beta^{by}_{xx'}.
\label{eq:inequality_inB}
\end{align}

Using these relations we can bound the Bell parameter. To this end, let $\ys$ be a choice for the value of $y$ made by Alice for a given $x$. Similarly, let $\xs$ be a choice for the value of $x$ made by Bob  for a given $y$. Then, 
\begin{equation}
	\begin{aligned}
		\mathcal{W}&=\somc E_{xy} \\
		&= \somc \som \a(x,y)\b(x,y) \\ 
		&\leq \somc \left(\som  \a(x,\tilde{y}_{x}) \b(x,y)+ \text{sign}(c_{xy})\sum_{a} \alpha^{ax}_{y\ys}) \right)\\
		&= \somc \som \a(x,\ys)\b(x,y) + \sum_{x,y} |c_{xy}| \sum_{a} \alpha^{ax}_{y\ys}\\
		&\leq \somc \left(\som   \a(x,\ys) \b(\xs,y)+ \text{sign}(c_{xy})\sum_{b} \beta^{by}_{x\xs}\right) +\sum_{xy} |c_{xy}| \sum_{a} \alpha^{ax}_{y\ys}, \\
		&=\somc \som \a(x,\ys)\b(\xs,y) +\sum_{x,y} |c_{xy}|  \left(\sum_{a} \alpha^{ax}_{y\ys}+\sum_{b}\beta^{by}_{x\xs}\right)\\
        & \leq {W}_{\text{LHV}} + \sum_{x,y} |c_{xy}|  \left(\sum_{a} \alpha^{ax}_{y\ys}+\sum_{b}\beta^{by}_{x\xs}\right),
	\end{aligned}
\end{equation}
where we have first used the SLHV model, in the third line we have used \eqref{eq:inequality_inA}, in the fifth line we have used \eqref{eq:inequality_inB} and in the last line we have identified the first term as the LHV bound. Note that because $\ys$ and $\xs$ are locally computed from the inputs of Alice and Bob respectively, their local outcomes are fully determined by their input and the hidden variable. Therefore, we recover the LHV bound in the last line. The upper bound is valid for every choice of $(\tilde{y}_0,\ldots,\tilde{y}_{m_B-1})$ and $(\tilde{x}_0,\ldots,\tilde{x}_{m_A-1})$. Therefore, the best bound corresponds to minimising over all possible choices of these tuples. Hence, 
\begin{equation}
	\mathcal{ W} \leq W_\text{LHV}+ \sum_{x,y} |c_{xy}| \min_{\tilde{x}_{y},\tilde{y}_{x}}\left(\sum_{a} \alpha^{ax}_{y\ys}+\sum_{b}\beta^{by}_{x\xs}\right).
\end{equation}
Note that for sufficiently large signalling parameters the right-hand-side will exceed the algebraic maxmimum of $\mathcal{W}$.

\section{From signalling LHS models to the geometrically inflated LHS set} \label{app:LHSgeom}
Here we show how Eqs.~(\ref{Eq:LHSsignalling}) and (\ref{SLHS}) lead to Eq.~(\ref{Eq:LHS_Sr}). By the dual formulation of the guessing probability, the allowed hidden states $\sigma_{\lambda,x}$ need to fulfill
$q_\lambda\sigma_{\lambda,x}\leq m_A W_\lambda$ with $\sum_\lambda \tr(W_\lambda)= \gamma$. Hence, one can write $q_\lambda\sigma_{\lambda,x} = m_A W_\lambda - (m_A W_\lambda-q_\lambda\sigma_{\lambda,x})$, where $m_A W_\lambda\geq 0$ and $m_A W_\lambda-q_\lambda\sigma_{\lambda,x}\geq 0$. Making these positive operators into states further gives $\sigma_{\lambda,x} = \frac{m_A \tr(W_\lambda)}{q_\lambda} \hat{W}_\lambda - \tr(A_{\lambda,x})\hat{A}_{\lambda,x}$, where $A_{\lambda,x}:=\frac{m_A}{q_\lambda}W_\lambda-\sigma_{\lambda,x}$. Writing $1+t:=\frac{m_A\tr(W_\lambda)}{q_\lambda}$, we get $\sigma_{\lambda,x}=(1+t)\hat{W}_\lambda - t\hat{A}_{\lambda,x}$. This shows that any assemblage described by Eq.~(\ref{Eq:LHSsignalling}) can be written as
\begin{equation}
\sigma_{a|x}=\sum_\lambda q_\lambda D^A_\lambda(a|x)\sigma_{\lambda,x}=\sum_\lambda q_\lambda D^A_\lambda(a|x)[(1+t)\hat{W}_\lambda - t\hat{A}_{\lambda,x}].
\end{equation}
This is the form given in Eq.~(\ref{Eq:LHS_Sr}) with the exception that we do not have yet a bound on the parameter $t$. To get a bound, we note that $t=\sum_\lambda q_\lambda t=\sum_\lambda q_\lambda\frac{m_A\tr(W_\lambda)}{q_\lambda}-1=m_A\sum_\lambda\tr(W_\lambda)= m_A\gamma$. This is exactly the guessing probability bound used in Eq.~(\ref{Eq:LHS_Sr}).

\section{Semi-definite programs for SLHS$_\gamma$ models} \label{app:sdpsteering}

Following the SDP formulation of the membership problem for the SLHS$_\gamma$ set, the robustness to that set can be $_\gamma$ written as
\begin{equation}
\begin{aligned}
    \text{Given} \quad & \gamma, \{\sigma_{a|x}\}_{a,x}\\
\min \quad & \mu \\
\text{s.t.} \quad & \frac{\sigma_{a|x}+ \mu \pi_{a|x}}{(1+\mu)} = \sigma_{a|x}^{\operatorname{SLHS}_\gamma}\\
& \sigma_{a|x}^{\operatorname{SLHS}_\gamma} = \sum_\lambda D(a|x,\lambda)\tilde{\sigma}_{\lambda, x}\\
& \sum_\lambda \tr(Z_\lambda) \leq \gamma\\
& Z_\lambda \succeq \frac{1}{m_A} \tilde{\sigma}_{\lambda,x} \quad \forall \lambda, x\\
& \tr(\tilde{\sigma}_{\lambda,x}) = p_\lambda \quad \forall \lambda, x \\
& \pi_{a|x} \succeq 0 \quad \forall a, x \quad
\sum_a\tr(\pi_{a|x}) = 1 \quad \forall x.
\end{aligned}
\end{equation}
Defining the new variables $\overline{\pi}_{a,x} = \mu\,\pi_{a|x}$, $\overline{\sigma}^{\operatorname{SLHS}_\gamma}_{a,x} = \sigma_{a|x}^{\operatorname{SLHS}_\gamma}(1+\mu)$, and $\overline{Z}_\lambda = Z_\lambda(1+\mu)$, as well as the unnormalized probabilities $\overline{p}_\lambda = p_\lambda(1+\mu)$, the problem leads to the following SDP:
\begin{align}
SR_{\text{SLHS}_\gamma}= \min_{\{\overline{\sigma}_{\lambda,x}, \overline{Z}_\lambda, \overline{p}_\lambda\}} \quad & \frac{1}{m_A} \sum_{a,x} \tr(\overline{\sigma}_{a|x}^{\operatorname{SLHS}_\gamma}) - 1 \\
\text{s.t.} \quad & \overline{\sigma}_{a|x}^{\operatorname{SLHS}_\gamma} - \sigma_{a|x} \succeq 0, \quad \forall a,x \label{st: succeq_pi}\\
& \overline{\sigma}_{a|x}^{\operatorname{SLHS}_\gamma} = \sum_\lambda D(a|x,\lambda) \overline{\sigma}_{\lambda,x}\\
& \sum_\lambda \tr(\overline{Z}_\lambda) \leq \gamma \frac{1}{m_A} \sum_{a,x} \tr(\overline{\sigma}_{a|x}^{\operatorname{SLHS}_\gamma}) \\
& \overline{Z}_\lambda \succeq \frac{1}{m_A} \overline{\sigma}_{\lambda,x}, \quad \forall \lambda, x \\
& \tr(\overline{\sigma}_{\lambda,x}) = \overline{p}_\lambda, \quad \forall \lambda, x  \\
& \overline{\sigma}_{\lambda,x} \succeq 0, \quad \overline{Z}_\lambda \succeq 0, \quad \overline{p}_\lambda \geq 0.
\end{align}
Positive values imply that the (possibly signalling) assemblage does not admit an SLHS$_\gamma$ model given $\gamma$. Furthermore, we can compute the critical robustness ($\min  \varepsilon\equiv SR_{\text{SLHS}_\gamma}^{\I}$) to white noise, such that $\frac{\sigma_{a|x}+\varepsilon \I/(d m_A)}{1+\varepsilon} = \sigma_{a|x}^{\operatorname{SLHS}_\gamma}$. This is achieved by replacing constraint \eqref{st: succeq_pi} with $\sigma_{a|x} + (\frac{1}{m_A} \sum_{a,x} \tr \overline{\sigma}_{a|x}^{\operatorname{SLHS}_\gamma}-1) \frac{\I}{(d m_A)} = \overline{\sigma}_{a|x}^{\operatorname{SLHS}_\gamma}$.

\section{Genuine high-dimensional steering with signalling}
\label{app:SN}
Below we present a simple examples of steering with signalling data.

In finite-dimensional systems, any unsteerable assemblage can be prepared from some separable state~\cite{Wiseman2007}. Hence, the presence of steering witnesses entanglement. This motivates one to ask what other properties of the state can be certified in a steering scenario. One natural candidate is the entanglement dimensionality of a bipartite state. In~\cite{designolle2021}, it was asked whether a given state assemblage can be prepared from a state with a bounded Schmidt number $n$. The Schmidt number is a quantifier of entanglement dimensionality: for a bipartite quantum state $\varrho_{AB}$ it is defined as
\begin{equation}
    SN(\varrho_{AB}) = \min_{\{p_\lambda,|\psi_\lambda\rangle\}} \max_\lambda \text{rank}(\text{tr}_A[|\psi_\lambda\rangle\langle\psi_\lambda|]),
\end{equation}
where the minimisation runs over all pure-state decompositions $\varrho_{AB} = \sum_\lambda p_\lambda |\psi_\lambda\rangle\langle\psi_\lambda|$. For two measurement settings $m_A = 2$, a linear steering witness constructed from two transposed mutually unbiased bases can be used for this task. The bound for a Schmidt-number-$n$ state is then given by
    $L_{SN\leq n} = \left(1+\frac{1}{\sqrt{d}}\right)\left(1+\frac{\sqrt{n}-1}{\sqrt{n}+1}\right)$,
as derived in~\cite{designolle2021}.

Similarly to the case of standard steering, one can make a signalling adjustment for Schmidt number detection. Following the procedure in Eqs.~(\ref{Eq:LinearInequalityLinearAdjustment}), one finds that the upper bound $L_{SN\leq n}$ can be linearly adjusted exactly as in the case of steering.

To give an example of Schmidt number detection from signalling data, we take the isotropic two-qutrit state $\varrho_{v}$ with visibility $v$, and let Alice measure in the computational basis, as well as in the Fourier basis. In the ideal case the (normalised) prepared states are of the form $\hat{\sigma}^{\text{NS}}_{a|x}= v \ket{\phi_{a|x}}\bra{\phi_{a|x}} + (1-v) \frac{\id}{3}$ where $\ket{\phi_{a|x}}$ are the projectors of Alice's measurements. As a theoretical model for signalling, we set $\sigma_{a|x} = \sqrt{\mu_x} \hat{\sigma}^{\text{NS}}_{a|x} \sqrt{\mu_x}$ where $\mu_1 = k \ket{0}\bra{0} + (1-k) \frac{\id}{3}$ and $\mu_2 = k \frac{(\ket{1}+\ket{2})(\bra{1}+\bra{2})}{2} + (1-k) \frac{\id}{3}$, with $k \in [0,1]$.

As Alice has only two measurement settings, the guessing probability can be calculated using the trace distance according to the Helstrom bound \cite{Helstrom1969}, i.e.
\begin{equation}\label{eq:Helstrom}
    P_g(\{\mu_1,\mu_2\})= \frac{1}{2}+\frac{1}{2} \norm{\frac{1}{2}\mu_1-\frac{1}{2}\mu_2}_1=\frac{1+k}{2} \:.
\end{equation}

\begin{figure}
    \centering
    \includegraphics[width=0.4\linewidth]{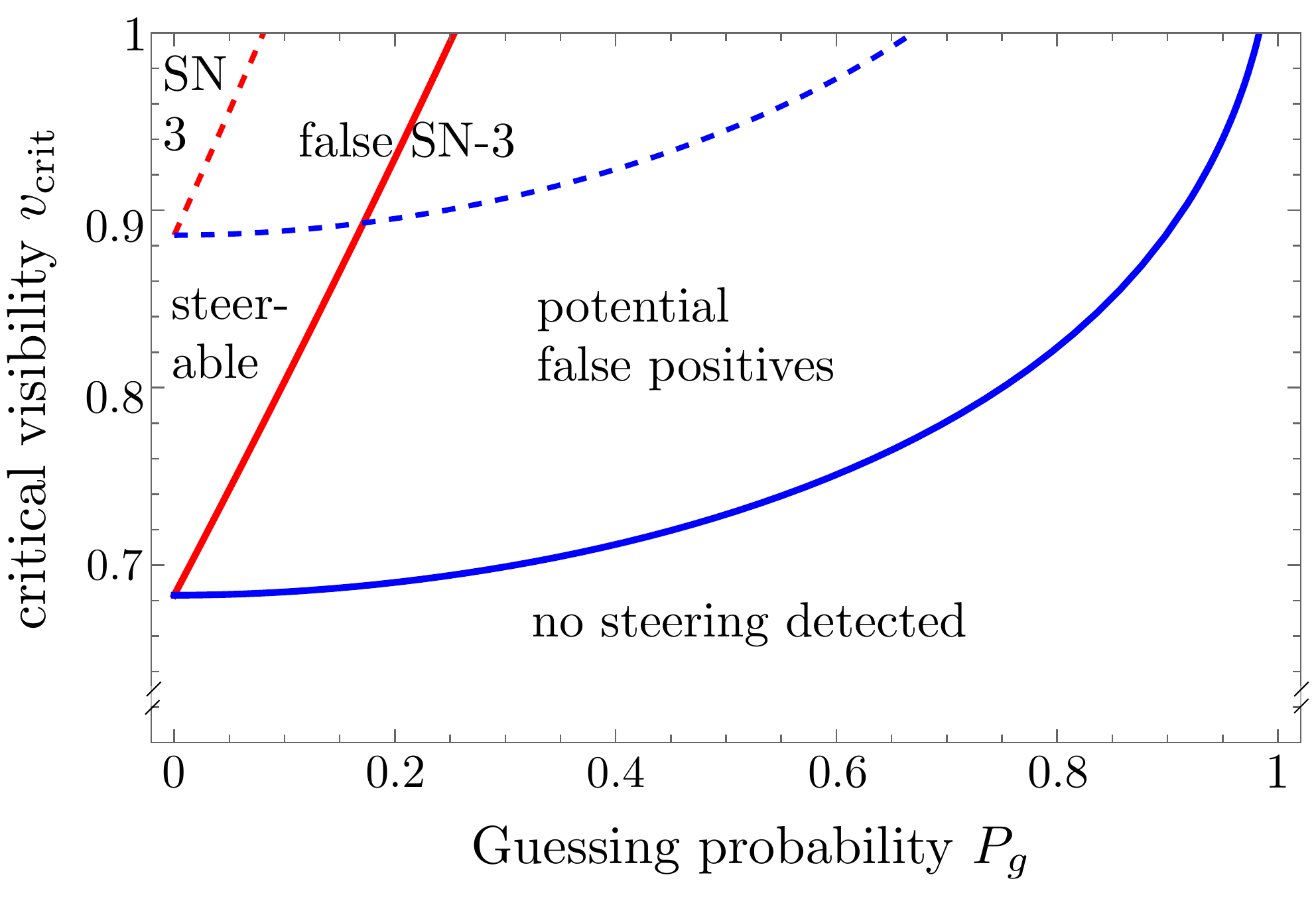}
    \caption{Critical visibility for witnessing high-dimensional steering. The solid lines represent the bounds for certifying entanglement, and the dashed lines represent the bounds for certifying Schmidt number $3$. Here, the blue and red represent the bounds obtained using the non-adjusted witness and the adjusted witness, respectively.}
    \label{fig:Exp_qutrit2input_Pg}
\end{figure}

By applying the presented linear witness to this example, the critical visibility necessary to certify entanglement dimensionality is determined, c.f. Fig.~\ref{fig:Exp_qutrit2input_Pg}. Notice that there is a region for which the non-adjusted witness certifies Schmidt number three, while the adjusted witness does not even certify entanglement.

\section{Signalling in post-selected  data} \label{app:postselection}
Consider that Alice's detector has an efficiency $\eta^A_{a\lvert x}\in[0,1]$ for outcome $a$ and input $x$. The efficiency can be estimated as 
\begin{equation}
	\eta^A_{a|x}= \frac{n_{x}(a)}{n_{x}(a)+n_{x}(\no)}, 
\end{equation}
where $n_{x}(a)$ is the number of events with outcome $a$ and input $x$ and $n_x(\no)$ is the number of no-click events. Since there is effectively one extra outcome the alphabet becomes $\tilde{a}\in[n_A]\cup \no$. We  similarly  have $\tilde{b}\in[n_A]\cup \no$ for Bob and associate detection efficiencies $\eta^B_{b\lvert y}$. The POVMs associated to the inefficient measurements become
\begin{align}
	A_{\tilde{a}|x} = \begin{cases}
		\eta_{a|x} A_{a|x} & \text{if } \tilde{a}=a, \\
		\openone - \sum_{a} \eta_{a|x} A_{a|x} &  \text{if } \tilde{a}=\no,
	\end{cases}
  &&	B_{\tilde{b}|y} = \begin{cases}
		\eta_{b|y} B_{b|y} & \text{if } \tilde{b}=b, \\
		\openone - \sum_{b} \eta_{b|y} B_{b|y} &  \text{if } \tilde{b}=\no
	\end{cases},
\end{align}
where  $\{ A_{a|x}\}$ and $\{ B_{b|y} \}$ are the POVMs of Alice and Bob with unit efficiency. 

Consider now that when Alice and Bob perform their measurements they post-select on the successful detection events, i.e.~when $\tilde{a}\in[n_A]$ and $\tilde{b}\in [n_B]$. The post-selected probability distribution becomes 
\begin{equation}
	p(a,b|x,y)=\frac{p(\tilde{a}=a ,\tilde{b}=b|x,y)}{N_{x,y}} = \frac{\eta_{a|x} \eta_{b|y}}{N_{x,y}}\tr( \left(A_{a|x}\otimes B_{b|y} \right)\rho),
\end{equation}
where we have defined $N_{x,y}=\sum_{\tilde{a} \neq \no , \tilde{b} \neq \no} p(\tilde{a},\tilde{b}|x,y)$ as the sum-total probability of both Alice and Bob obtaining a successful event.

For Alice, signalling can be tested as follows:

\begin{equation}
	\begin{aligned}
		&p(a|x,y)-p(a|x,y')=\sum_{b } p(a,b|x,y)-p(a,b|x,y')=\eta_{a|x} \tr(\left( A_{a|x} \otimes (F_{x,y}-F_{x,y'}) \right) \rho),
	\end{aligned}
\end{equation}
where we have defined the operator $F_{x,y}$ as
\begin{equation}
	F_{x,y}= \frac{1}{N_{x,y}}\sum_{b} \eta_{b|y} B_{b|y}.
\end{equation}
For no-signalling to hold we require that $F_{x,y}-F_{x,y'}=0$. This is satisfied  when $\eta_{b\lvert y}^B=\eta^B$. Analogously,  no-signalling from Bob to Alice holds if $\eta_{a\lvert x}^A=\eta^A$. These restrictions mean that $N_{x,y}=\eta^A\eta^B$ and hence $F_{x,y}=\frac{\openone}{\eta^A}$ and similarly for the case of signalling to Alice.  However, if the efficiencies depend on the outcome or input, then  $F_{x,y}-F_{x,y'}\neq 0$ and no-signalling is violated.

In the main text we considered the situation in which the efficiencies depend only on the  outcome  irrespective of the setting and the party, i.e.~ $\eta^A_{k\lvert z}=\eta^B_{k\lvert z}=\eta_k$. We consider the standard maximal violation of the CHSH inequality, which is associated with choosing $\rho$ as the maximally entangled two-qubit state, and letting Alice and Bob perform local anticommuting Pauli measurements. It leads to the probability distribution $p^{\text{ideal}}(a,b|x,y)=\frac{1}{4}\left(1+(-1)^{a+b+xy}/\sqrt{2}\right)$ which achieves $\mathcal{W}_\text{CHSH}=2\sqrt{2}$. Next, we consider that the measurements are  implemented with the outcome-dependent inefficient detectors. This leads to 
\begin{equation}
\begin{aligned}
    N_{x,y}&= \sum_{\tilde{a},\neq \no ,\tilde{b} \neq \no } p(\tilde{a},\tilde{b}|x,y) \\
    &=\sum_{a,b} \eta_{a} \eta_{b} p^{\text{ideal}}(a,b|x,y) = \eta_{0}^2 p^{\text{ideal}}(0,0|x,y)+ \eta_{0} \eta_{1} \left(p^{\text{ideal}}(0,1|x,y)+p^{\text{ideal}}(1,0|x,y)\right) + \eta_{1}^2 p^{\text{ideal}}(1,1|x,y)\\
    &=\frac{1}{4}\eta_{0}^2 \left(1+(-1)^{xy}/\sqrt{2}\right)+\frac{1}{4} \eta_{0}\eta_{1} \left(2- 2 (-1)^{xy}/\sqrt{2}\right) + \frac{1}{4}\eta_{1}^2 \left( 1+ (-1)^{xy} /\sqrt{2}\right)\\
    &=\frac{1}{4}\left( (\eta_{0}+\eta_{1})^{2} + (-1)^{xy} (\eta_{0}-\eta_{1})^2/\sqrt{2}\right).
\end{aligned}
\label{eq:maximalchshs_normalization}
\end{equation}

\begin{figure}
    \centering
    \includegraphics[width=0.55\linewidth]{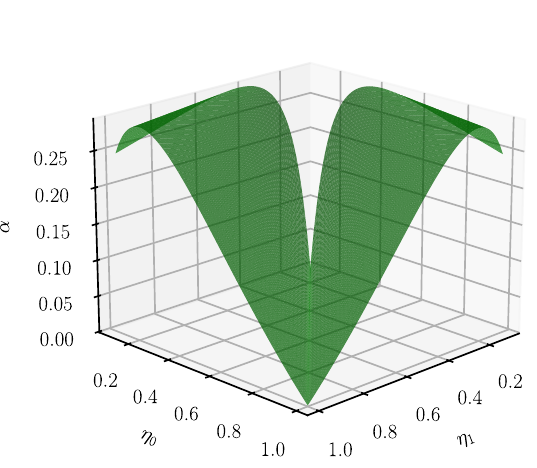}
    \caption{Signalling parameter $\alpha$ plotted as a function of efficiencies $\eta_{0}$ and $\eta_{1}$ as described in \ref{eq:Alpha_eta_function}.}
    \label{fig:alpha_eta_plot}
\end{figure}

The apparent signalling can now be calculated
 \begin{equation}
 \begin{aligned}
    \alpha^{ax}_{yy'}&=|p(a|x,y)-p(a|x,y')|= \left |\sum_{b}p(a,b|x,y)-p(a,b|x,y') \right|\\
    &=\left|\sum_{b}\frac{\eta_{a}\eta_{b}}{4} \left( \frac{1+(-1)^{a+b+xy}/\sqrt{2}}{N_{xy}}- \frac{1+(-1)^{a+b+xy'}/\sqrt{2}}{N_{xy'}}\right) \right|, \\
    &=\left|\frac{\eta_{a}}{4} \left(\frac{\eta_{0}+\eta_{1}+(-1)^{a+xy}(\eta_{0}-\eta_{1})/\sqrt{2}}{N_{xy}}-\frac{\eta_{0}+\eta_{1}+(-1)^{a+xy'}(\eta_{0}-\eta_{1})/\sqrt{2}}{N_{xy'}}\right) \right|,\\
    &= \eta_{a}\left|\frac{\eta_{0}+\eta_{1}+(-1)^{a+xy}(\eta_{0}-\eta_{1})/\sqrt{2}}{ (\eta_{0}+\eta_{1})^{2} + (-1)^{xy} (\eta_{0}-\eta_{1})^2/\sqrt{2}}-\frac{\eta_{0}+\eta_{1}+(-1)^{a+xy'}(\eta_{0}-\eta_{1})/\sqrt{2}}{ (\eta_{0}+\eta_{1})^{2} + (-1)^{xy'} (\eta_{0}-\eta_{1})^2/\sqrt{2}}\right|.
    \end{aligned}
 \end{equation}
This is non-zero only when $x=1$. Also, one can verify that the above value is independent of $a$. This means that it suffices to consider the signalling parameter $\alpha=\alpha^{11}_{01}$. By analogy, when considering signalling from Bob to Alice the only relevant parameter becomes $\beta=\beta^{11}_{01}$. Since we have selected the efficiencies to depend only on the outcome it follows that $\alpha=\beta$. Its value is 
\begin{equation}
    \alpha= \eta_{1}\left|\frac{\eta_{0}+\eta_{1}+(\eta_{0}-\eta_{1})/\sqrt{2}}{\left( (\eta_{0}+\eta_{1})^{2} - (\eta_{0}-\eta_{1})^2/\sqrt{2}\right)}-\frac{\eta_{0}+\eta_{1}-(\eta_{0}-\eta_{1})/\sqrt{2}}{\left( (\eta_{0}+\eta_{1})^{2} +  (\eta_{0}-\eta_{1})^2/\sqrt{2}\right)}\right|.
    \label{eq:Alpha_eta_function}
\end{equation}
This signalling parameter $\alpha$ is plotted in Fig.~\ref{fig:alpha_eta_plot}.

Post-selection of the local behaviour $p_{\text{local}}(a,b|x,y)=\frac{1}{4} (1 + (-1)^{a+b+xy})$ results in a similar signalling parameter  if $\sqrt{2}$ is replaced with $2$ in (\ref{eq:Alpha_eta_function}). Interestingly, if the efficiencies are equal, the behaviour achieves the local maximum of the CHSH inequality at $2$. Once asymmetric efficiencies are allowed the behaviour attains larger values $W_{\text{CHSH}}\geq 2$. In a standard Bell test one would, falsely, conclude that the behaviour is nonlocal. 

\begin{figure}
    \centering
    \includegraphics[width=0.5\linewidth]{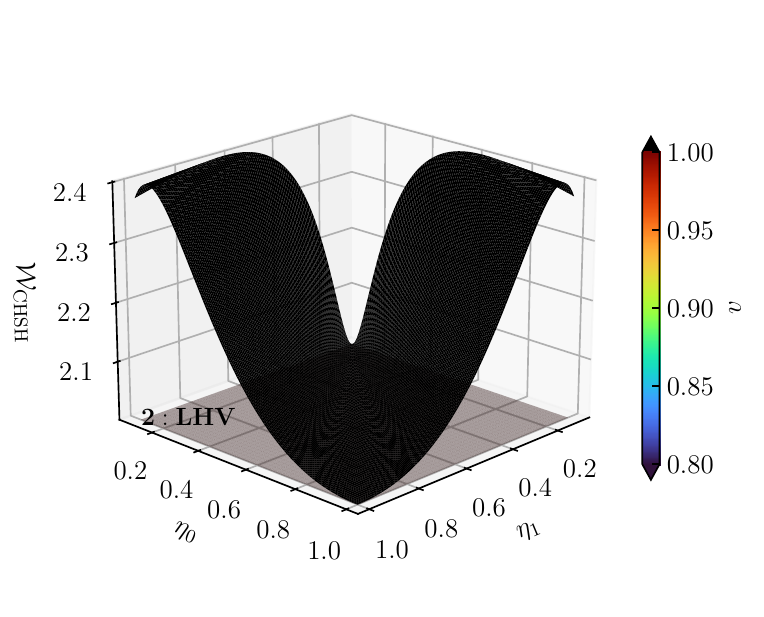}
    \caption{The CHSH parameter plotted as a function of detection efficinceis $\eta_{0}$ and $\eta_{1}$ of the post-selection of the behaviour $p(a,b|x,y)=\frac{1}{4}\left(1+(-1)^{a+b+xy}\right)$. The heat map represents the visibility of the post-selectd behaviour in the SLHV model with $\alpha^{ax}_{yy'}=|p(a|x,y)-p(a|x,y')|$ and $\beta^{by}_{xx'}=|p(b|x,y)-p(b|x',y)|$.} 
    \label{fig:CHSH_surface_LHV}
\end{figure}

In Fig.~\ref{fig:CHSH_surface_LHV}, we plot the CHSH parameter for the post-selected behaviour $p_{\text{local}}$ as a function the efficiencies $\eta_{0}$ and $\eta_{1}$. The distance to the SLHV model, indicated as a heat map, is characterized with the white noise visibility $v$ using the LP provided in \ref{section:Visibility_dual}. The signalling parameters for the SLHV model are calculated from the post-slected local behaviour $p'_{\text{local}}$ as $\alpha^{ax}_{yy'}=|p_{\text{local}}'(a|x,y)-p_{\text{local}}'(a|x,y')|$ and similarly for $\beta^{by}_{xx'}$. Notably, the visibility remains constant and equal to one despite increasing the asymmetry in the efficiencies and thus the apparent signalling. The plot demonstrates how the SLHV model can prevent false positive, that arise due to apparent signalling, in nonlocality tests.

\end{document}